\documentclass[twocolumn,preprintnumbers,superscriptaddress,prl,nofootinbib,longbibliography,amsmath,amssymb]{revtex4-1}

\usepackage{graphicx}
\usepackage{dcolumn}
\usepackage{bm}
\usepackage{color}
\usepackage{amsmath}
\usepackage[version=3]{mhchem} 
\usepackage{xcolor}
\usepackage[T1]{fontenc}   
\usepackage{lmodern}       
\usepackage[utf8]{inputenc}
\usepackage[unicode]{hyperref}

\usepackage{epstopdf}
\usepackage{epsfig}

\begin{document}

\title  {Structural transition and possible pressure-induced superconductivity in a suboxide La$_5$Pb$_3$O}

\author{Jiaqiang Yan}
\affiliation{Materials Science and Technology Division, Oak Ridge National Laboratory, Oak Ridge, TN 37831, United States}

\author{David Singh}
\affiliation{Department of Physics and Astronomy, University of Missouri, Columbia, MO, 65211, United States}

\author{Bayram Saparov}
\affiliation{Department of Chemistry and Biochemistry, Stephenson Life Sciences Research Center, University of Oklahoma, Norman, OK 73019, United States}

\author{Huibo Cao}
\affiliation{Neutron Scattering Division, Oak Ridge National Laboratory, Oak Ridge, TN 37831, United States}

\author{Yejun Feng}
\affiliation{Okinawa Institute of Science and Technology Graduate University, Onna, Okinawa 904-0495, Japan}

\author{Jinguang Cheng}
\affiliation{Institute for Solid State Physics, University of Tokyo, Kashiwa, Chiba 277-8581, Japan}

\author{Yoshia Uwatoko}
\affiliation{Institute for Solid State Physics, University of Tokyo, Kashiwa, Chiba 277-8581, Japan}

\author{David Mandrus}
\affiliation{Materials Science and Technology Division, Oak Ridge National Laboratory, Oak Ridge, TN 37831, United States}
\affiliation{Department of Materials Science and Engineering, University of Tennessee, Knoxville, Tennessee 37996, USA}

\date{\today}

\begin{abstract}
Here we report a structural phase transition and its possible competition with superconductivity in the suboxide La$_5$Pb$_3$O. Upon cooling through $T_t$ = 225 K, La$_5$Pb$_3$O transforms from a high-temperature \textit{I}4/\textit{mcm} to a low-temperature \textit{P}4/\textit{ncc} structure in which La–Pb dimerization along the c-axis occurs. This transition is accompanied by anomalies in the temperature dependence of electrical resistivity and specific heat. High-pressure electrical transport measurements reveal that hydrostatic pressure suppresses the structural transition and possibly induces superconductivity with a maximum superconducting temperature of 10 K. Density functional theory calculations show minimal changes in the electronic density of states and no gap opening at $E_F$ across $T_t$, suggesting that the transition is driven by bonding effects rather than Fermi surface instability. These findings establish La$_5$Pb$_3$O as a promising platform for exploring the interplay between weak structural transitions and superconductivity.
  
\end{abstract}

\maketitle

\section{Introduction}

Suboxides form a distinct class of oxides in which the electropositive element is present in excess compared to conventional stoichiometric oxides. Therefore, suboxides are often referred to as "metal-rich" compounds.  This unusual stoichiometry often results in extensive metal–metal bonding that gives rise to intricate metal cluster arrangements and unusual physical and chemical properties \cite{simon1997group,nohara2024metal}. In suboxides with exceptionally high metal-to-oxygen ratios, oxygen atoms frequently occupy interstitial positions within the metal framework. The intentional incorporation of small atoms, such as carbon, boron, nitrogen, fluorine, or oxygen, into these interstitial sites can stabilize otherwise unattainable phases, as exemplified by the discovery of the Nd$_2$Fe$_{14}$B permanent magnet \cite{croat2022history}. Even without altering the overall crystal framework, such void-filling light elements can profoundly modify the physical properties, for example, the inclusion of oxygen can increase the superconducting temperature of Nb$_5$Ir$_3$O and the mechanical properties of Ti$_4$Co$_2$O superconductors\cite{ma2021group, lifen2025synergetic}.

The hexagonal Mn$_5$Si$_3$-type (D8$_8$) crystal structure, belonging to the $P$6$_3$/$mcm$ space group, is notable for its ability to accommodate foreign atoms within its interstitial sites, thereby forming "filled" Mn$_5$Si$_3$-type compounds \cite{corbett1998widespread,corbett2000exploratory}. These interstitial sites can be occupied by a wide range of elements. For example, seventeen La$_5$Pb$_3Z$ ($Z$=B, C, P, S, Cl, As, Se, Sb, Cr, Mn, Fe, Co, Ni, Cu, Zn, Ru, and Ag) derivatives of La$_5$Pb$_3$ with the stuffed Mn$_5$Si$_3$-type structure were synthesized by Guloy and Corbett three decades ago\cite{guloy1994exploration}. Interestingly, incorporation of O or N into the lattice stabilizes the tetragonal Cr$_5$B$_3$-type structure ($I$4/$mcm$) \cite{guloy1992synthesis}. To date, the physical properties of these structural variants have not been thoroughly studied yet.

In this work, we investigate the structure evolution of La$_5$Pb$_3$O through x-ray and neutron single crystal diffraction, specific heat, and electrical resistivity measurements on millimeter-sized single crystals. Below 225 K, La and Pb atoms dimerize along the c-axis resulting in anomalies in both resistivity and specific heat. Transport measurements under hydrostatic pressure reveal that pressure suppresses this transition and likely induces superconductivity at low temperatures. Density functional theory calculations indicate that the transition is driven by bonding effects rather than Fermi surface instability.

\section{Experimental details}
\subsection{Crystal growth}
La$_5$Pb$_3$O single crystals were grown out of a mixture of La:Co:Pb in a ratio of 7:2:1 following the recipe that Macaluso et al. developed for the growth of Ce$_5$Pb$_3$O\cite{macaluso2004structure}. The formation of La$_5$Pb$_3$O is facilitated by the reaction between Al$_2$O$_3$ crucible and La melt\cite{yan2015flux}. In this growth, Co is critical in lowering the melting temperature of the melt but will not contaminate the crystals; Al$_2$O$_3$ crucible oxidizes La forming a passivating layer of La$_2$O$_3$ which provides oxygen to the melt and protects La from further oxidization. In a typical growth,  La (99.999\% Ames Laboratory), Co (99.999\%, Alfa Aesar), and Pb (99.9999\%, Alfa Aesar) were placed in a 2 ml alumina growth crucible  and sealed in a quartz tube together with an alumina catch crucible of the same dimension containing quartz wool. The sealed ampoule was heated to 1150$^o$C in 8 hours. After dwelling at 1150$^o$C for 4 hours, it was then cooled to 850$^o$C over 72 hours. A faster cooling rate may lead to the precipitation of La$_5$Pb$_3$. Once the furnace reached 850$^o$C, the excess flux was decanted from the crystals. This growth yields rectangular La$_5$Pb$_3$O  crystals with the typical dimension of 0.5$\times$ 0.5$\times$ 2-4mm$^3$. 

\subsection{X-ray powder and single crystal diffraction}
Room temperature X-ray powder diffraction patterns were collected on a X'Pert PRO MPD X-ray Powder Diffractometer using the Ni-filtered Cu-K radiation.  Data collections were carried out in 10-90$^o$ 2$\theta$ range with a step size of 0.02$^o$  and a counting time of 120 seconds/step in a continuous scan mode. The phase identification was carried out employing X'Pert HighScore Plus software.

Single crystal X-ray diffraction measurements were carried out on a Bruker SMART APEX CCD-based single crystal X-ray diffractometer. Selected under a microscope, a rectangular needle crystal was cut to a suitable size (0.059$\times$0.044$\times$0.036 mm$^3$) inside Paratone N oil. The X-ray intensity data were collected at 173(2)K and 273(2)K with Mo K$\alpha$ ($\lambda$ = 0.71073  {\AA} ) radiation. The structure was solved by direct methods and refined by full matrix least-squares methods on F2 using the SHELXTL software package. Absorption correction was applied using SADABS within the SHELXTL package.

\subsection{Synchrotron and neutron single crystal diffraction}
Synchrotron x-ray diffraction measurements were performed at Sector 4-ID-D of the Advanced Photon Source, Argonne National Laboratory, to study the structural transition and look for reflections associated with possible incommensurate charge density wave. For synchrotron x-ray diffraction measurements, a bar shaped sample of 1.5$\times$0.29$\times$0.25 mm$^3$ size was attached to a copper stick at one end using GE varnish and further sealed inside a Be dome filled with helium exchange gas. The whole assembly was performed inside a protective Ar atmosphere.  The longitudinal direction of the sample is parallel to the c-axis and the four naturally grown side surfaces are normal to [1, 1, 0]. The sample mosaic varied between 0.02 - 0.04 degrees FWHM. Search for incommensurate charge density wave state was carried out using hard x-ray diffraction at Sector 4-ID-D. The majority of diffraction measurement was performed using 20.0 KeV x-rays in the reflection (Bragg) geometry. The x-ray beam was focused to a spot size of 0.25 (H) $\times$ 0.1 (V) mm$^2$ by a pair of Pd mirrors. In addition, unfocused x-rays of 35.0 KeV energy with a spot size of 0.6 (H) $\times$ 0.55 (V) mm$^2$ were also used in the transmission (Laue) diffraction geometry, which render our sample thickness to about two absorption lengths and allow for full bulk sensitivity.

Single crystal neutron diffraction was measured at HB-3A four-circle diffractometer at the High Flux Isotope Reactor at the Oak Ridge National Laboratory. One piece of crystal with the dimensions of 0.6$\times$0.8$\times$2.6 mm$^3$ and mass of 14 mg was used in the measurement. The data were collected at 300 K with a neutron wavelength of 1.5424 {\AA} from a bent perfect Si-220 monochromator. The measurement at room temperature was aimed to determine the oxygen content by using the large neutron cross section of oxygen. The temperature dependence measurements were to track the structure evolution. 

\subsection{Electrical resistivity and specific heat Measurements}
Electrical resistivity measurements were performed using a standard four-probe technique in a Quantum Design Physical Property Measurement System (QD PPMS) in the temperature range 1.9$\leq$T$\leq$300 K.  A rectangular crystal with typical dimension of 0.4$\times$0.6$\times$3 mm$^3$ was selected for the measurements. Silver paste was used to create contacts and Pt wires were placed such that the current is applied along the crystallographic c-axis. The specific heat data were collected using the heat capacity option of a QD PPMS  in the temperature interval 1.9$\leq$T$\leq$300 K.

Electrical resistivity measurements under high pressure were performed using a four-probe method with a cubic anvil apparatus \cite{cheng2018cubic}, which can generate quasihydrostatic pressures of up to 8 GPa. A preheated pyrophyllite cube was used as the gasket and glycerol as the pressure transmitting medium. Each set of measurements was carried out at a fixed pressure while warming up the sample from the lowest temperature.

\section{Experimental Results}
\subsection{Crystal stoichiometry and stability}
Room temperature x-ray powder diffraction of pulverized single crystals confirmed single phase of as-grown crystals. As mentioned earlier, Co is an important component in forming a proper flux mediating the growth of La$_5$Pb$_3$O, but it doesn not contaminate the as-grown crystals. This is confirmed by our elemental analysis and magnetic measurements. No sign of Co contamination is observed by either technique.

\begin{figure} \centering \includegraphics [width = 0.47\textwidth] {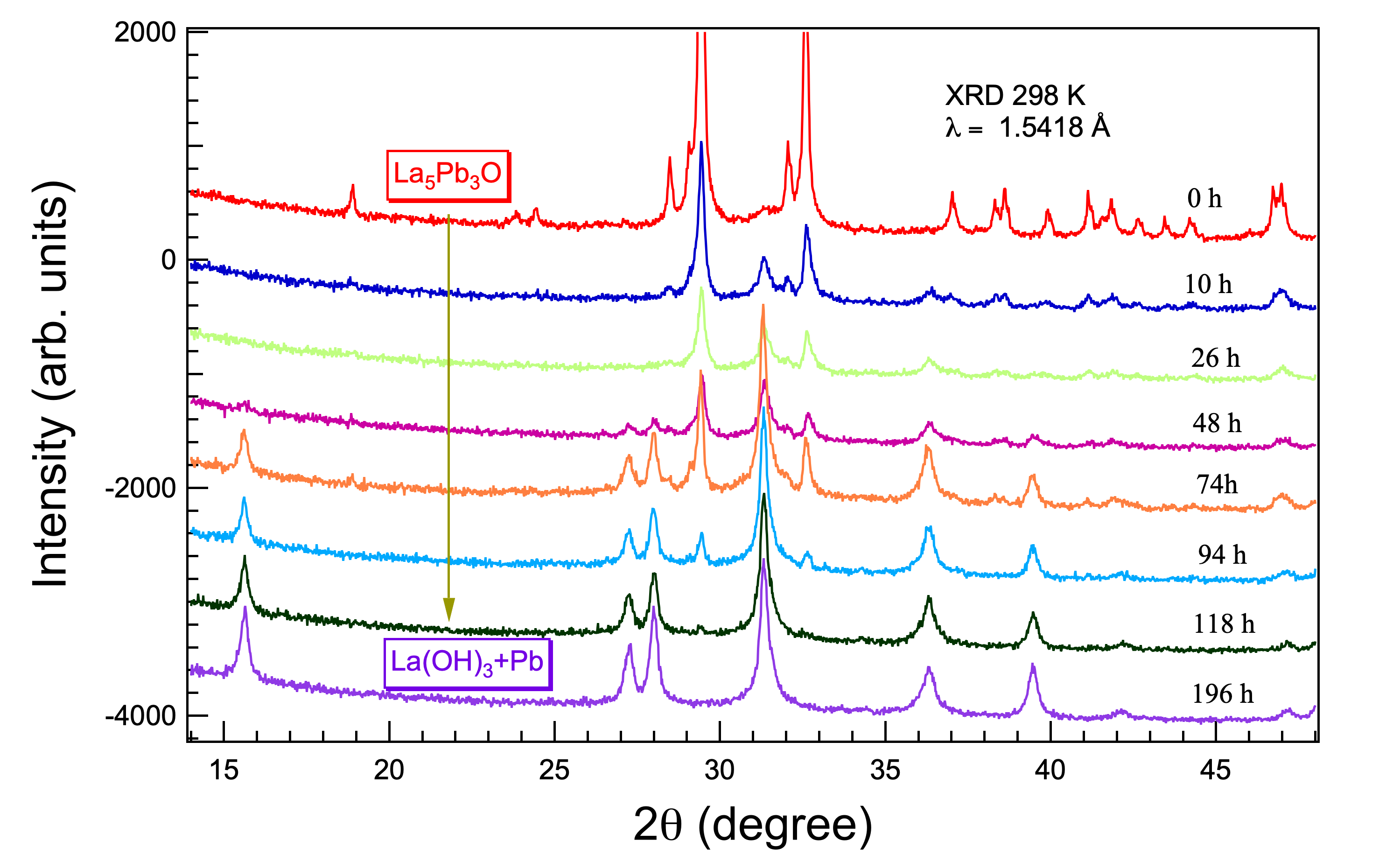}
\caption{(color online) Evolution with time of X-ray powder diffraction patterns of pulverized La$_5$Pb$_3$O crystals exposed in air at room temperature. The patterns are shifted vertically for clarity. }
\label{XRD-1}
\end{figure}

The shining La$_5$Pb$_3$O crystals are stable in air for a couple of days. However, the pulverized single crystals are rather air sensitive. As shown in Fig. \ref{XRD-1}, after exposing in air for 10 hours, the intensity of reflections from La$_5$Pb$_3$O drops significantly: those weak reflections around 2$\theta$\,=\,24 degree disappear; the intensity of the strongest reflection around 2$\theta$\,=\,29 degree drops by 70\%. After exposing in air for 5 days, La$_5$Pb$_3$O decomposes into La(OH)$_3$ and Pb. To avoid the crystal degradation, as-grown crystals were kept in a dry glove box filled with helium. As-grown crystals were used in most measurements to avoid any polishing or cutting. The exposure of crystals in air was controlled to be as short as possible in all the measurements performed in this study.

We think that in our growth at high temperatures the alumina crucible oxidized La metal in the melt and forms a La$_2$O$_3$ shell which serves as the source of oxygen to the melt and a passivation layer to prevent further oxidization of La metal \cite{yan2015flux}. Based on the source of oxygen, it is reasonable to question the oxygen content in the as-grown crystals. Since oxygen has a larger cross section to neutrons than x-rays, single crystal neutron diffraction was performed to determine the oxygen content. Refinement of the neutron diffraction data collected at room temperature gives an oxygen site occupancy of 0.90(3), indicating slight oxygen deficiency. This oxygen deficiency suggests that the oxygen stoichiometry in La$_5$Pb$_3$O can be tuned while preserving the Cr$_5$B$_3$-type structure. By varying the cooling rate during crystal growth, we obtained La$_5$Pb$_3$O crystals with the structure transition temperature varying in the range of 180-240\,K. We assume this variation in transition temperature results from different oxygen content. While the current work focuses on crystals with the structure transition at T$_t$=225\,K, the impact of oxygen nonstoichiometry on the structural transition and physical properties deserves further study.

\begin{figure} \centering \includegraphics [width = 0.47\textwidth] {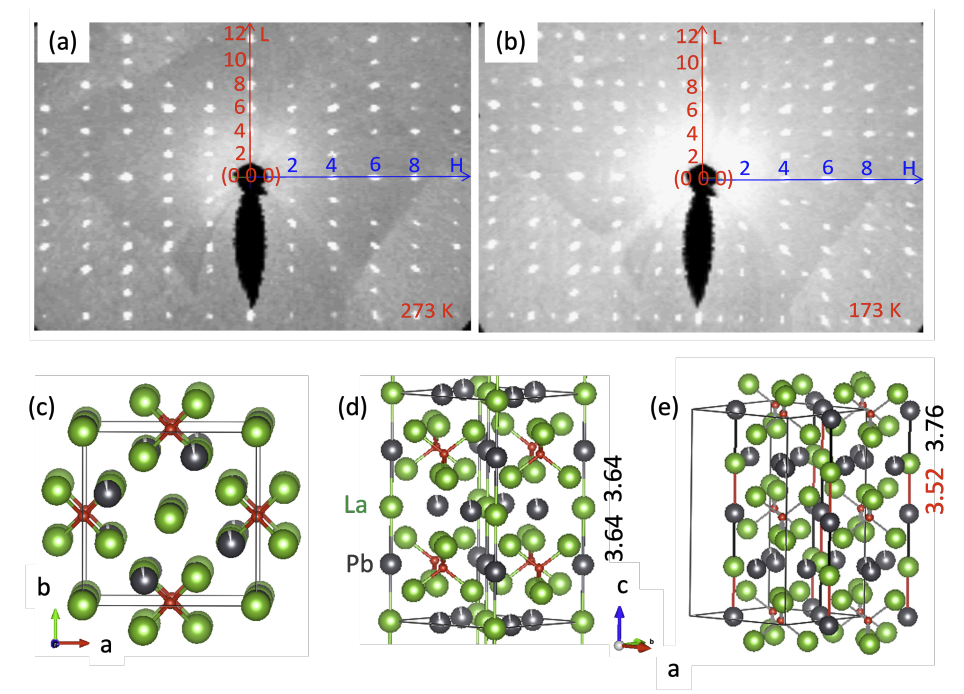}
\caption{(color online) Structure of La$_5$Pb$_3$O. (HOL) cut at (a) 273K and (b) 173K. (c) View along \textit{c}-axis to highlight that oxygen stays in the center of La tetrahedra. (d) High temperature structure (\textit{I}4/\textit{mcm}) with uniform La-Pb bond of 3.64$\AA$ along the \textit{c}-axis. (e) Below the structure transition, one La shifts toward Pb along the \textit{c}-axis leading to La-Pb dimerization in the low temperature structure (space group \textit{P}4/\textit{ncc}). The shorter La–Pb bond measures 3.52$\AA$, compared to 3.76$\AA$ for the longer one.}
\label{structure-1}
\end{figure}

\subsection{Single crystal diffraction}
The atomic coordinates of Ce$_5$Pb$_3$O \cite{macaluso2004structure} were used in the refinements of our single crystal diffraciton data collected at 273\,K and all sites were refined anisotropically. The resultant R-values were elevated with R1$\approx$0.035 and wR2$\approx$0.109 (note here that these R-values are comparable to those reported for Ce$_5$Pb$_3$O \cite{macaluso2004structure}), and the difference Fourier map showed uneven features with highest peak/hole $\approx$4.86/-6.36 e$^-/\AA^3$. Consequently, site occupation factors (SOF) were checked by freeing individual occupancy factors of atom sites. This procedure revealed SOF of $\approx$0.95 on Pb2 site and resulted in improved R-values as well as difference Fourier map.

The same structure information was initially used to refine the data collected at 173K. The refinement gives reasonable R factors. However, detailed investigation of the data reveals that about 1600 weak reflections are missing. Figure\,\ref{structure-1} (a) and (b) show the (H0L) cut of the data taken at 273K and 173K, respectively. Superlattice peaks are clearly observed at low temperatures, and all can be indexed with the $P$4/$ncc$ space group, indicating a structural phase transition occurring between 173 K and 273 K. The refinement results are summarized in the Supplemental Material. 

In the structure of La$_5$Pb$_3$O, oxygen stays in the center of the tetrahedra formed by four La atoms. This is highlighted in Fig.\,\ref{structure-1}(c). These tetrahedra interpenetrate with La centered octahedra creating a three dimensional framework. These structure features remain the same in the whole temperature range. At low temperatures, one La moves toward Pb along the crystallographic c-axis leading to La-Pb dimerization and periodic lattice distortion. This change is highlighted in Fig.\,\ref{structure-1}(d,e). The low temperature phase still holds the inversion symmetry. So this structure transition is different from that in LiOsO$_3$ where a ferroelectric like structure transition occurs in a metal\cite{shi2013ferroelectric}. 

The structure transition determined here agrees well with a most recent thorough investigation of the structure of this compound\cite{penacchio2025charge}. The structure transition occurs at a lower temperature in our crystals possibly due to the oxygen nonstoichiometry.

\begin{figure} \centering \includegraphics [width = 0.47\textwidth] {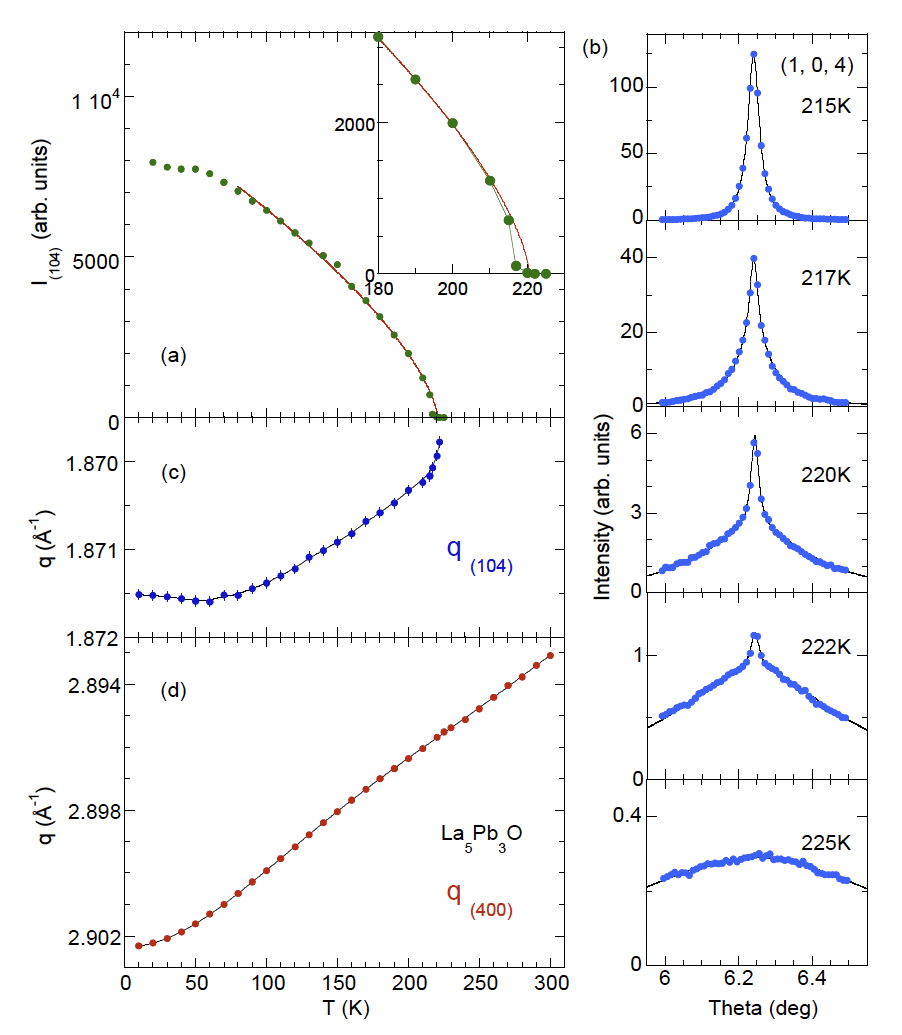}
\caption{(color online) Synchrotron x-ray study of phase transition in La$_5$Pb$_3$O. (a) Diffraction intensity of the (1, 0, 4) reflection that is forbidden in the high temperature  phase shows strong temperature dependence below T$_t$. For 0.35 T$_t<$\,T\,$<$\,T$_t$, the evolution could be fit to a power law I=(T$_t$-T)$^2\beta$ (solid red) with $\beta$ = 0.337(11).  (inset) Close to T$_t$, the intensity evolution shows a weak first order discontinuity. The data was measured upon warming. (b) The rocking curves of the sample at various temperatures show the presence of both a sharp mosaic peak and a broad diffusive intensity. Both profiles could be fit with a Lorentzian shape with different widths (black line). Only the 217, 220, and 222K data require fitting to a summation of both profiles. (c, d) The measured lattice wave vector as a function of temperature. We notice for (4, 0, 0) order, there is no anomaly across T$_t$, while for (1, 0, 4) order, the wave vector shows a singularity at the phase boundary. }
\label{Synchrotron-1}
\end{figure}

\subsection{Synchrotron and neutron diffraction}

In addition to determine the oxygen occupancy, we also used neutron diffraction to monitor the structure transition. Figure S1 in the Supplemental Material shows the temperature dependence of the intensity of (104) reflection which is forbidden for the high temperature phase. The intensity change clearly shows that the structural transition takes place at T$_t$=225K.

To further resolve the details of the structural transition and look for weak and possibly incommensurate superlattice reflections, we performed synchrotron x-ray single crystal diffraction. As shown in Fig. \ref{Synchrotron-1}(a), diffraction intensities of forbidden orders such as (1,0,4) in the high temperature tetragonal phase show up below T$_t$. This agrees with the neutron single crystal diffraction results shown in Fig. S1. Our results also reveal the change of lattice structure upon cooling across T$_t$. In Fig. \ref{Synchrotron-1}(b), we show the progressive development of sample rocking curve at various temperatures near T$_t$. Over a narrow temperature range, the sharp component that is indicative of a long-range order and the broad component which is due to diffusive scattering of fluctuating order, exist simultaneously. The diffraction intensity measured from the sharp mosaic peak is plotted in Fig. \ref{Synchrotron-1}(b), showing a weak first order transition at T$_t$. On the other hand, over a large temperature range, the gradual rise of diffraction intensity with the temperature follows a power law behavior of \textit{I} $\sim$ (T$_t$-T)$^{2\beta}$. In the scenario of atomic displacements, the diffraction intensity \textit{I} is typically proportional to the lattice distortion $\varepsilon$ as \textit{I} $\sim$ $\varepsilon$ $^2$. We notice the measured critical exponent 2$\beta$ = 0.674$\pm$0.023 is close to the exponent $\gamma$  measured from the excess resistivity $\bigtriangleup$ $\rho$ $\sim$ (T$_t$-T)$^\gamma$. These results suggest close correlation between the structure transition and transport anomaly.

Scanning along major reciprocal lattice directions such as (H, 0, 0), (H, H, 0), and (0, 0, L) revealed no trace of incommensurate charge density wave. Along (H, 0, 0) and (H, H, 0), we achieved an elastic scattering background level at 4.4 x 10$^{-8}$ and 1.2 x 10$^{-7}$ of the (4, 0, 0) and (3, 3, 0) intensities respectively. The (0, 0, L) direction was probed using 35 KeV x-rays with a background level at 2.3 x 10$^{-7}$ of (0, 0, 6) peak intensity. The statistics of background at around 1000 counts per reciprocal lattice point would allow detection of incommensurate charge density wave states further down to an intensity level at 10\% of the background. Nevertheless, no incommensurate charge density wave state was observed. Additional searches along (0.5, 0.5, L) and (H, H, 0.5) yielded no evidence of any incommensurate charge density wave either.

\subsection{Specific heat and electrical resistivity under ambient pressure}

\begin{figure} \centering \includegraphics [width = 0.47\textwidth] {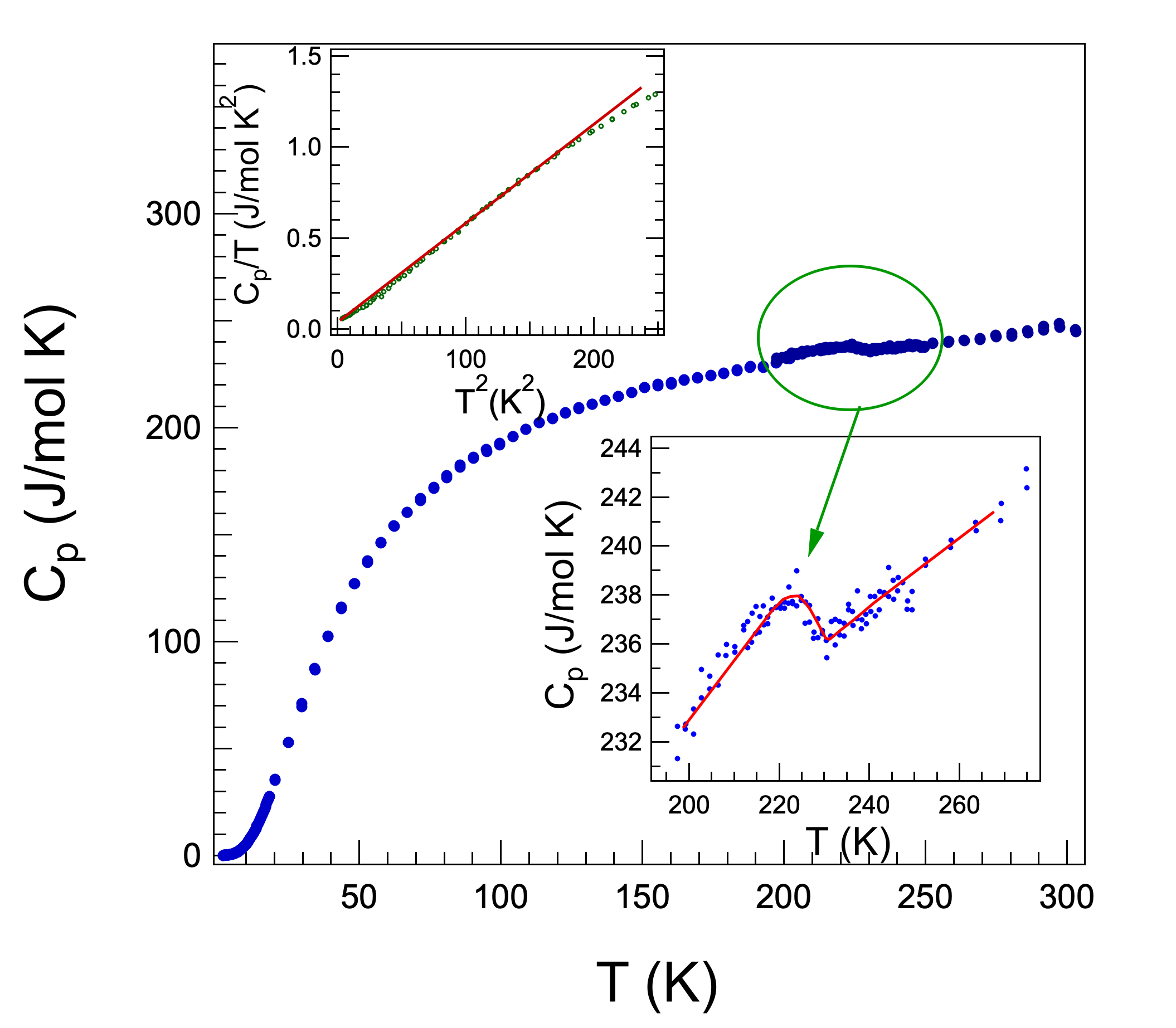}
\caption{(color online) Temperature dependence of specific heat. The upper inset shows the linear fitting of Cp/T vs T$^2$ in the low temperature regime. The lower inset highlights the weak anomaly around the structure transition. The solid curve is a guide to the eyes.}
\label{Cp-1}
\end{figure}

Figure \ref{Cp-1} shows the temperature dependence of specific heat measured in the temperature range 1.9K$\leq$T$\leq$300K. The heat capacity at room temperature attains a value of 246J/mol K, which is little larger than the classical high temperature Dulong-Petit value of 3nR=225J/mol K at constant volume, where R is the molar gas constant and n=9 is the number of atoms per formula unit. The upper inset of Fig. 7 shows the fitting of low temperature specific heat data  to C$_p$ = $\gamma$ $T+$ $\beta$T$^3$. The fitting yields $\gamma$= 25(2) mJ/mol K$^2$ and $\beta$= 0.55mJ/mol K$^4$. The Debye temperature for La$_5$Pb$_3$O is 316\,K. The lower inset highlights the details around 220K. The detailed measurement reveals a weak lambda-like anomaly starting at 230K. The transition temperature agrees with that observed from resistivity and diffraction measurements.

\begin{figure} \centering \includegraphics [width = 0.47\textwidth] {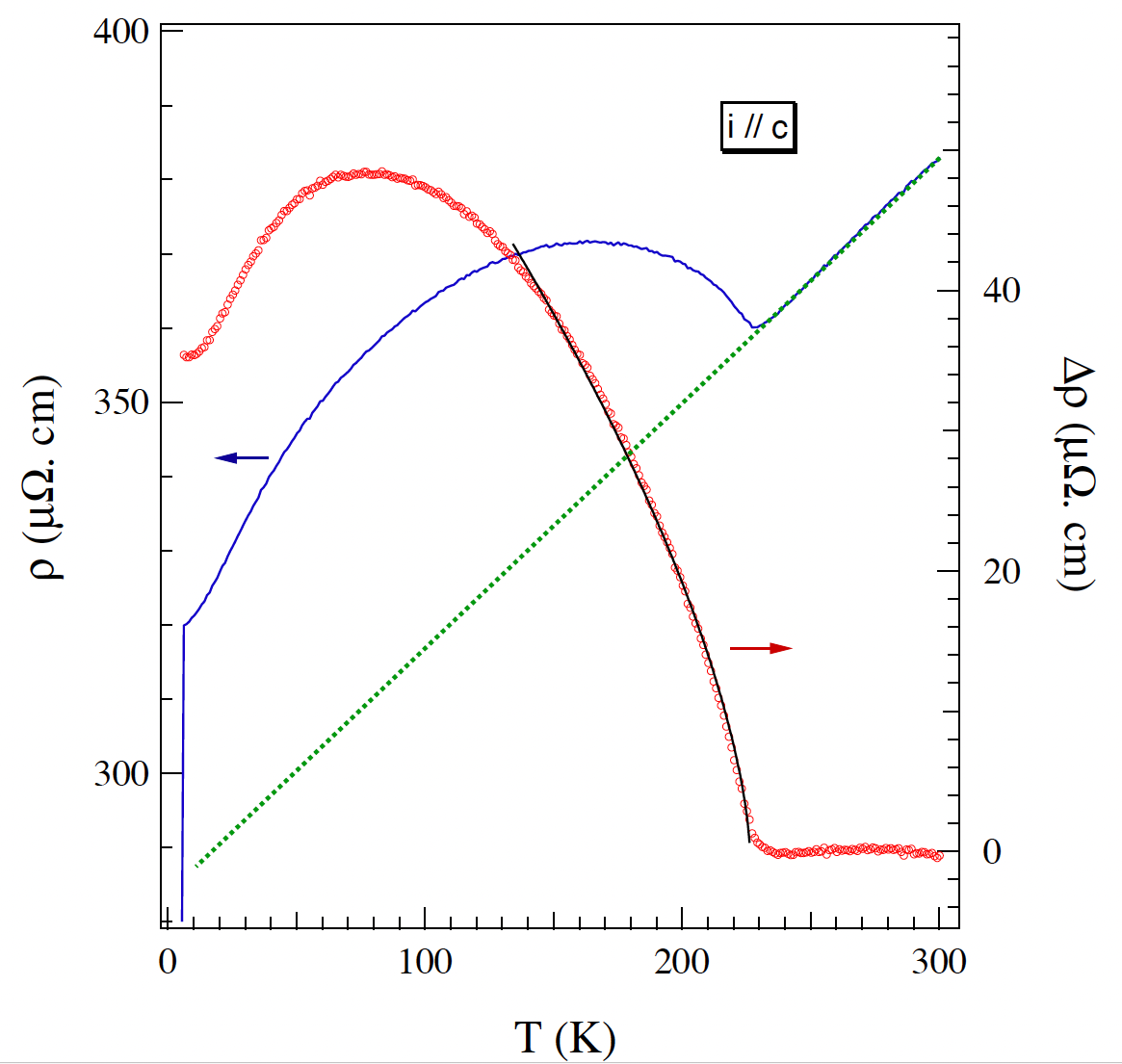}
\caption{(color online) Temperature dependence of electrical resistivity measured with current parallel to the crystallographic c-axis. The dashed line (green) shows the linear extrapolation of the high temperature resistivity. The residual resistivity (red, open circle) is obtained by subtracting the linear extrapolation from the low temperature resistivity.}
\label{RT-1}
\end{figure}

The structure transition is also reflected in the temperature dependence of electrical resistivity. Figure \ref{RT-1} shows the temperature dependence of the electrical resistivity measured in the temperature range 2K$\leq$T$\leq$300K with current parallel to the crystallographic \textit{c} axis. The sharp drop of electrical resistivity at about 7 K comes from the superconducting transition of Pb on the crystal surface. Above T$_t$=225 K, the electrical resistivity shows a linear temperature dependence. Below 225 K, the electrical resistivity increases with decreasing temperature and reaches a maximum at about 150 K. The residual resistivity, obtained by subtracting the linear extrapolation of the high temperature resistivity, can better highlight the transport response to the structure transition when cooling below T$_t$=225 K. As described below, the resistivity anomaly becomes weaker when measured with the electrical current in ab-plane and when measured under high pressure. The residual resistivity can help track the evolution with high pressure of the structure transition. The temperature dependence of the electrical resistivity was measured in both warming and cooling processes and no hysteresis was observed in the temperature range studied. The measurements in an applied magnetic field up to 140 kOe observe no field dependence.

\subsection{Electrical resistivity under high pressure}
\begin{figure} \centering \includegraphics [width = 0.47\textwidth] {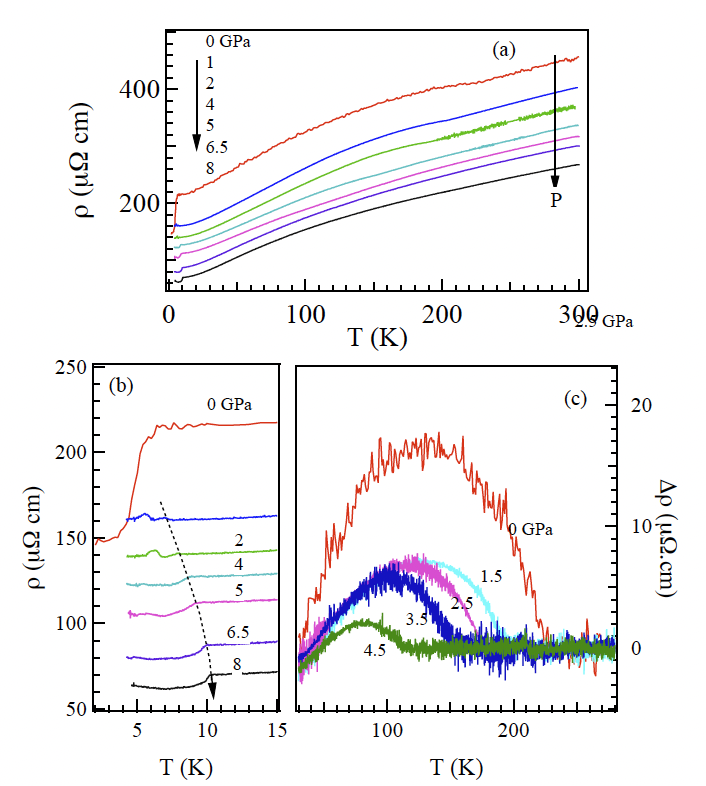}
\caption{(color online) (a) The temperature dependence of electrical resistivity under high pressure measured with  i//ab. (b) The resistivity drop at low temperatures due to the pressure induced superconductivity. (c) the temperature dependence of residual resistivity after subtracting the high temperature linear extrapolation. }
\label{HPRT-1}
\end{figure}

Figure \ref{HPRT-1} shows the temperature dependence of electrical resistivity under pressures up to 8\,GPa with the electrical current in ab plane. At ambient pressure, the resistivity enhancement cooling across the structural transition is smaller in magnitude compared to that when measured with electrical current parallel to the crystallographic c-axis. This is supported by the theoretical analysis described later. With increasing pressure, the resistivity enhancement shifts to lower temperatures and becomes smaller in magnitude. Above 5\,GPa, the resistivity enhancement is barely observable. The above pressure dependence of the resistivity anomaly across T$_t$ is better illustrated by plotting the residual resistivity after subtracting the high temperature linear extrapolation. As shown in Fig.\,\ref{HPRT-1}(c), the structure transition occurs at lower temperatures under high pressure. At the pressure of 4.5\,GPa, a weak anomaly around 120\,K is observed. 

Under ambient pressure, a drop in resistivity around 7\,K is also observed, which is due to the filamentary superconductivity of lead from flux. The pressure dependence of superconducting temperature, T$_c$, of lead has been well studied and T$_c$ is suppressed under high pressures. With increasing pressure, the resistivity anomaly due to lead is suppressed to lower temperatures. Above 4\,GPa, one new resistivity drop above 7\,K can be well resolved(see Fig.\,\ref{HPRT-1}b). The resistivity drop takes place at higher temperatures with increasing pressure. This resistivity drop and its positive pressure dependence were observed in multiple pieces of samples. The only variation between samples is the magnitude of the resistivity drop. Unfortunately, no zero resistivity was observed in all samples with a well defined structure transition around 200\,K. While other explanations for the observed resistivity drop cannot be excluded, it is likely a sign of filamentary superconductivity induced by high pressure. The crystal quality may prevent the observation of zero resistivity. When cutting the crystals to smaller pieces to fit the pressure chamber, we noticed that some white colored inclusions can be observed in the crystals. Considering the formation of La$_2$O$_3$ shell during the crystal growth\cite{yan2015flux}, it is not a surprise to have some La$_2$O$_3$ inclusions in the crystals. Recently, Penacchio et al.\cite{penacchio2025charge} grew La$_5$Pb$_3$O single crystals out of Ta tube with controlled oxygen content in the starting materials and obtained better crystals. It will be very interesting to study under high pressure the transport properties of these high quality single crystals. 

\begin{figure} \centering \includegraphics [width = 0.47\textwidth] {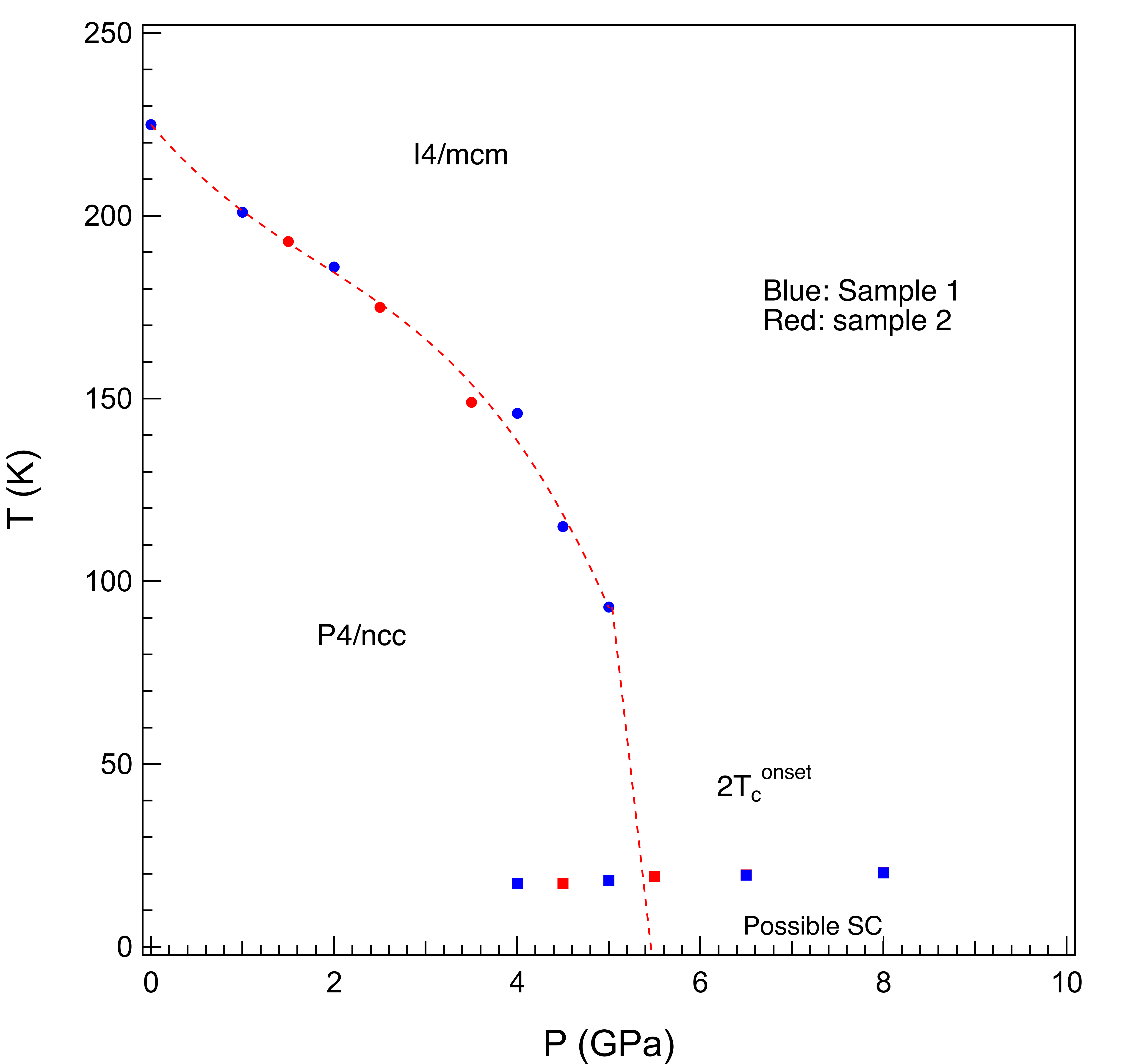}
\caption{(color online) Phase diagram under high pressure. Since no zero resistivity is observed, we labeled the dome between 4 and 6\,GPa as possible superconductivity. }
\label{PD-1}
\end{figure}

Figure\,\ref{PD-1} shows the proposed T-P phase diagram that illustrates the competition of the structure transition and possible low temperature superconductivity. As shown in Fig. S3 in Supplemental Material, a similar suppression of the structure transition is also observed for Ce$_5$Pb$_3$O. Again, no zero resistance but a drop in resistivity was observed at high pressures possibly due to the oxide contamination because Ce$_5$Pb$_3$O crystals were grown using the same approach. However, these results suggest that $R_5$Pb$_3$O ($R$=rare earth) can serve as a new material platform for the study of the interplay between structure transition and superconductivity.  Studies to even higher temperatures are needed. In addition, high pressure x-ray diffraction studies are desired to look for possible structure transition under high pressure.

\section{Electronic structure}
We did density functional theory calculations for both the high- and low-T phases based on the experimentally determined crystal structures (at 273 K and 173 K, respectively). These (PBE-GGA) \cite{perdew1996generalized} functional, including spin orbit coupling (SOC) and the general potential linearized augmented planewave (LAPW) method\cite{singh2006planewaves} as implemented in the WIEN2k code\cite{blaha2020wien2k}. The LAPW method is an all-electron method, that does not involve pseudopotentials or frozen core approximations. We used LAPW sphere radii of 2.5 bohr for La and Pb and 1.9 bohr for O, with well converged basis sets including local orbitals for the semicore states and dense samplings of the Brillouin zone. These amounted to 32x32x32 for the high temperature phase and 32x32x24 for the low temperature phase. These dense meshes allowed precise determination of the Fermi level and the density of states.

\begin{figure} \centering \includegraphics [width = 0.47\textwidth] {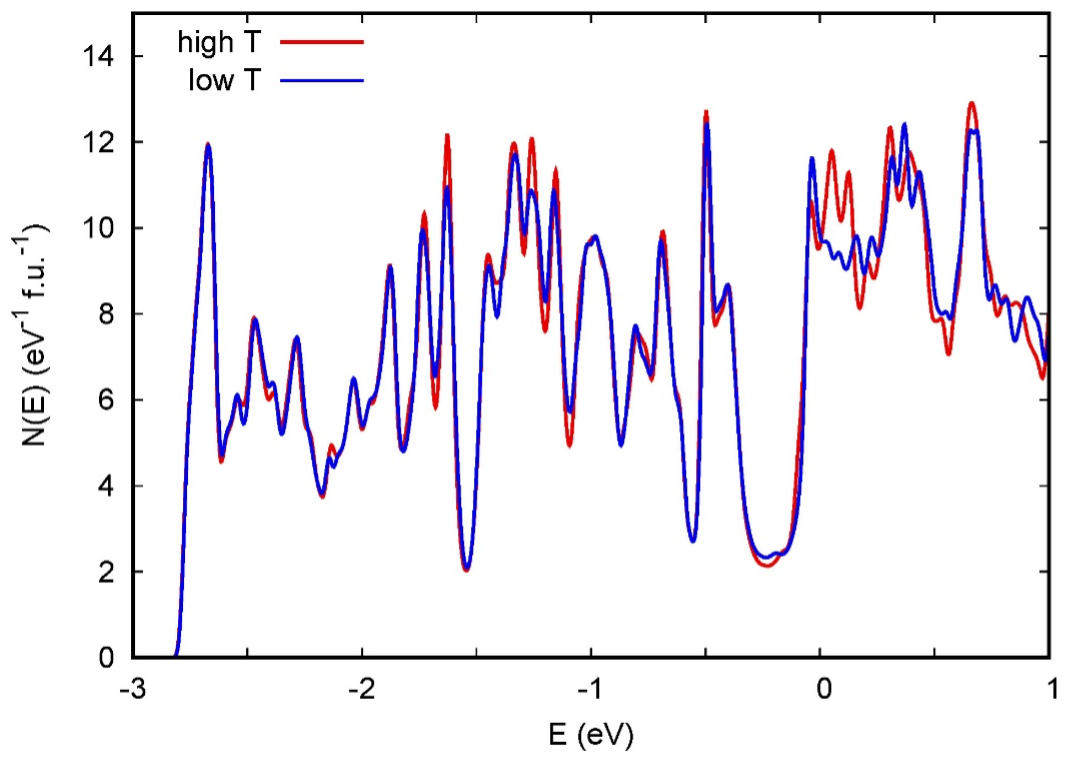}
\caption{(color online) Comparison of the electronic DOS of the high-T and low-T structures of La$_5$Pb$_3$O. The DOS is shown on a per formula unit basis with the Fermi level at 0 eV. }
\label{DOS-1}
\end{figure}

A comparison of the total electronic density of states (DOS) of the high-T and low-T phases is given in Fig.\,\ref{DOS-1}. As seen, there are only very small changes in the DOS in the valence band region. The result near the Fermi level, E$_F$, is similar to that of Penacchio and co-workers \cite{penacchio2025charge}, in that there is a small dip in the DOS of the low temperature phase approximately 0.1 eV above E$_F$. However, this dip is much less pronounced in our calculation. In any case, the dip is not at E$_F$ and the DOS at E$_F$, N(E$_F$) is higher in the low temperature structure. Specifically, N(E$_F$)=9.4\,eV$^{-1}$ and N(E$_F$)=9.5\,eV$^{-1}$ for the high temperature and low temperature structures, respectively. The small changes in the DOS, and the lack of gapping at E$_F$ differ from the standard picture of a charge density wave, which is driven by reconstruction of the electronic structure at E$_F$.

In order to better characterize the electronic structure changes at E$_F$, we calculated the plasma frequences using the optical package of the WIEN2k code. We obtain $\hbar\Omega_{p,xx}$=1.48\,eV and $\hbar\Omega_{p,zz}$=1.20\,eV for the high temperature structure and $\hbar\Omega_{p,xx}$=1.32\,eV and $\hbar\Omega_{p,zz}$=0.97\,eV for the low temperature structure. We cross checked these values by 
calculated the conductivity over relaxation time ($\sigma/\tau$) with the BoltzTraP code \cite{madsen2006boltztrap} (note that $\sigma/\tau$ is proportional to ($\hbar\Omega_{p,xx}$)$^2$).  Thus, the plasma frequency is significantly reduced in the low temperature structure. Thus, a 
transport signature of the transition is to be expected, for example in resistivity, particularly for transport along the c-axis (z direction). Note also that while the relaxation time $\tau$ normally depends on temperature and also may differ between the two structures, a change in $\tau$ would not directly affect the anisotropy. The conductivity anisotropy is predicted to be $\sigma_z$/$\sigma_x$=0.66 for the high temperature phase and $\sigma_z$/$\sigma_x$=0.54 for the low temperature phase, providing a transport signature of the transition independent of the values of the relaxation time.

\begin{figure} \centering \includegraphics [width = 0.47\textwidth] {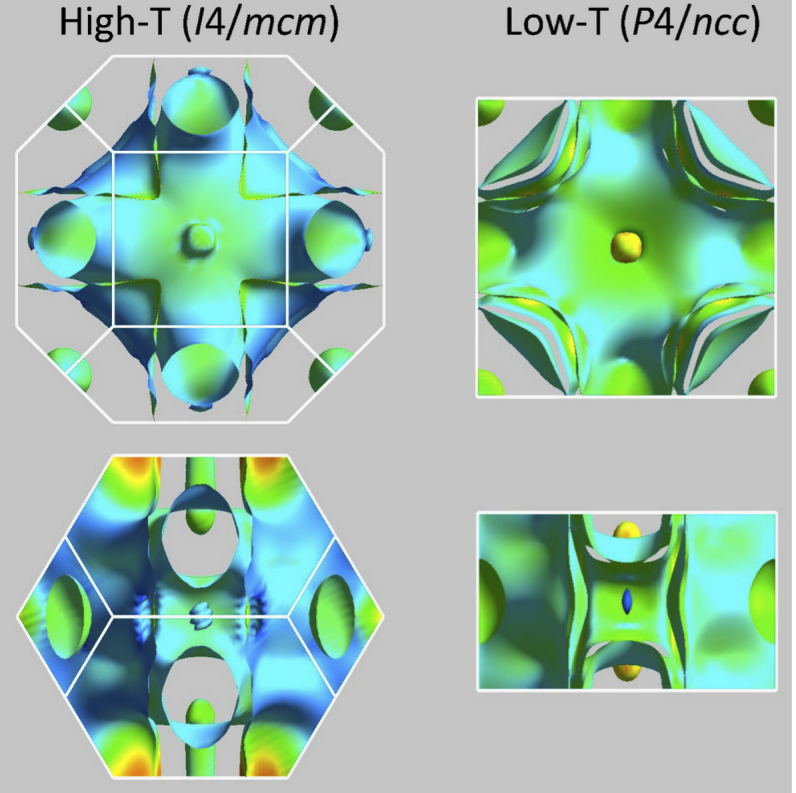}
\caption{(color online) Fermi surfaces of the high (left) and low temperature (right) phases. The top panels show views along the kz (c-axis) direction, while the bottom are perpendicular to it. The shading is by Fermi velocity, with blue representing low velocity and red representing high velocity.}
\label{Fermi-1}
\end{figure}

The structure of the low temperature phase consists of a doubling of the high temperature primitive unit cell, going from a body centered cell to a simple tetragonal cell with similar lattice parameters. Thus, the Brillouin zone is folded correspondingly through the transition, folding the Fermi surface as well. The high temperature structure has four sheets of Fermi surface: a small hole pocket and small electron sheet, and two large open sheets. The low temperature phase has seven sheets of Fermi surface. These are a small hole sheet similar to the high temperature phase, two small electron sheets and four open sheets. The Fermi surfaces are shown in Fig.\,\ref{Fermi-1}, and the individual sheets are shown in Fig. S4 in Supplemental Material. It is to be noted that the Fermi surface of the high temperature phase shows some flat sections, but that these sections persist in the low temperature phase, meaning that they are not gapped by the structural transition. This is indicative of a structural transition driven by bonding effects, for example, mismatched atomic sizes rather than a charge density wave based on a Fermi surface instability.

\section{Summary}
In summary, we study the structure transition of the suboxide La$_5$Pb$_3$O and find its low temperature phase features the La-Pb dimerization. Hydrostatic pressure gradually suppresses this transition and it vanishes near 5\,GPa. As this structure  transition disappears, the resistivity exhibits a drop that we attributed to pressure-induced superconductivity that onsets around 4\,GPa and reaches $T_c\approx$10\,K at 8\,GPa, the highest pressure studied in this work. Preliminary data on Ce$_5$Pb$_3$O show a similar correlation between the structural transition and superconductivity under high pressure, suggesting $R_5$Pb$_3$O ($R$ = rare earth) as a platform for studying the interplay between structural instabilities and superconductivity.

We did not observe zero resistance under pressure likely due to oxide inclusions in the crystals. Further high-pressure transport, magnetic, and structural measurements on cleaner crystals are needed to confirm the low temperature superconductivity and its competition with the structure transition. It would also be valuable to partially substitute O with F or N to tune band filling and examine the corresponding response of both the structural transition and superconductivity. Further research is also needed to determine if La$_5$Pb$_3$N undergoes a similar structural transition and, if so, exhibits a similar competition between the structure transition and superconductivity under high pressure.

\section{Acknowledgment}
Work at ORNL was supported by U.S. Department of Energy,  Office of Science, Basic Energy Sciences, Materials Sciences and Engineering Division. This research at ORNL’s High Flux Isotope Reactor was sponsored by the Scientific User Facilities Division, Office of Basic Energy Sciences, US Department of Energy.

%

\end{document}